\documentclass[fleqn,10pt]{wlscirep}
\usepackage[utf8]{inputenc}
\usepackage[T1]{fontenc}

\usepackage{multirow}
\usepackage{listings}
\usepackage{xcolor}
\usepackage{colortbl}
\usepackage{color,soul}
\usepackage{todonotes}
\usepackage{bm}
\usepackage{amsmath}
\usepackage{url}

\newcommand{\TrainingSet}{\bm{T}} 
\newcommand{\TestSet}{\Psi}

\title{The OPS-SAT benchmark for detecting anomalies in satellite telemetry}

\author[1,3,*]{Bogdan Ruszczak}
\author[3]{Krzysztof Kotowski}
\author[4]{David Evans}
\author[2,3]{Jakub Nalepa}
\affil[1]{Faculty of Electrical Engineering, Automatic Control and Informatics, Department of Informatics, Opole University of Technology, Prószkowska Str. 76, 45-758 Opole, Poland}
\affil[2]{Faculty of Automatic Control, Electronics and Computer Science, Department of Algorithmics and Software, Silesian University of Technology, Akademicka Str. 16, 44-100 Gliwice, Poland}
\affil[3]{KP Labs, Bojkowska Str. 37J, 44-100 Gliwice, Poland}
\affil[4]{European Space Agency/ESOC, Robert-Bosch-Str. 5, 64293 Darmstadt, Germany}

\affil[*]{corresponding author(s): Bogdan Ruszczak (b.ruszczak@po.edu.pl)}

\begin{abstract}

Detecting anomalous events in satellite telemetry is a critical task in space operations. This task, however, is extremely time-consuming, error-prone and human dependent, thus automated data-driven anomaly detection algorithms have been emerging at a steady pace. However, there are no publicly available datasets of real satellite telemetry accompanied with the ground-truth annotations that could be used to train and verify anomaly detection supervised models. In this article, we address this research gap and introduce the AI-ready benchmark dataset (OPSSAT-AD) containing the telemetry data acquired on board OPS-SAT---a CubeSat mission which has been operated by the European Space Agency which has come to an end during the night of 22--23 May 2024 (CEST). The dataset is accompanied with the baseline results obtained using 30 supervised and unsupervised classic and deep machine learning algorithms for anomaly detection. They were trained and validated using the training-test dataset split introduced in this work, and we present a suggested set of quality metrics which should be always calculated to confront the new algorithms for anomaly detection while exploiting OPSSAT-AD. We believe that this work may become an important step toward building a fair, reproducible and objective validation procedure that can be used to quantify the capabilities of the emerging anomaly detection techniques in an unbiased and fully transparent way.

\end{abstract}
\begin{document}

\flushbottom
\maketitle

\thispagestyle{empty}



\section*{Background \& Summary}


The anomaly detection (AD) domain encompasses a diverse array of methodologies for the identification of anomalous patterns in data of various modalities. These approaches can be applied to a multitude of data types, including images, text, and time series data, among others. However, the development and evaluation of real-world anomaly detection applications are dependent on the availability of real-world data. Currently, there is a considerable number of datasets available for a wide range of scenarios~\cite{pang2021deep}, but the satellite telemetry data for AD is an extremely underrepresented category in this catalogue. This kind of data is difficult and costly to obtain, often confidential, and requires expert knowledge to annotate properly. The only two widely accessible and used collections of this type include the NASA Soil Moisture Active Passive (SMAP) and Mars Science Laboratory (MSL) datasets~\cite{hundman_detecting_2018}. They offer short fragments of signals and related commands from 55 and 27 telemetry parameters, respectively, with a total of 105 annotated anomalies. However, the recent consensus in the community is that they should not be used for time series AD benchmarking due to their unrealistic anomaly density, many trivial anomalies, mislabelled ground truth, distributional shifts, and a lack of meaningful correlation between commands and channels \cite{wu_current_2023, wagner_timesead_2023, amin_maleki_sadr_anomaly_2023}. Other well-known satellite telemetry datasets, such as Mars Express \cite{Petkovic2022} or NASA WebTCAD \cite{sanchez_webtcad_2018}, do not contain annotations of anomalous events. There is an ongoing activity to publish a large-scale AD dataset by European Space Agency (ESA) solving all the mentioned issues~\cite{kotowski23, kotowski_european_2024}, but it will primarily address the needs of large-scale, complex and relatively stable missions. 

The dataset introduced in this article, dubbed OPSSAT-AD, is fundamentally different from those available in the literature, as it tackles a very specific ESA OPS-SAT mission---a CubeSat flying laboratory, for which we might expect a noticeable number of abnormal events~\cite{evans_ops-sat_2016}. The raw telemetry from OPS-SAT is characterized by many data gaps, artifacts, sampling frequency changes, and signal amplitude variations. The dataset was collectively curated by space operations engineers and machine learning experts to make it useful for building and validating data-driven anomaly detection techniques. It includes a selection and the corresponding ground-truth annotation of 2123 short single-channel satellite telemetry fragments (univariate time series) captured within 9 telemetry channels. Due to the underlying nature of the OPS-SAT mission, anomalous fragments account for 20\% of the dataset. Such fragments contain raw data with many aforementioned real-life challenges, and they differ in their length and sampling frequency. For each telemetry fragment, the dataset also contains a set of 18 handcrafted features used in the actual machine learning AD algorithm validated on board OPS-SAT~\cite{OPSiccs23}. These features are exploited in this article to benchmark 30 other supervised and unsupervised machine learning algorithms for anomaly detection. All of them were trained on 1494 and tested on 529 telemetry segments, and assessed using 7 metrics suggested for quantifying the operational capabilities of anomaly detection algorithms---this training-test dataset split is included in our benchmark as well.

Overall, the benchmark (including the dataset, training-test dataset split, suggested quality metrics, and our baseline results) introduced in this paper shall help the community to create and compare their approaches to detecting anomalies in real-life satellite telemetry in a fair and unbiased way. Therefore, we also address the reproducibility crisis currently observed in the (not only) machine learning community~\cite{KAPOOR2023100804}. While the OPS-SAT spacecraft completed its atmospheric reentry at the end of May 2024, its successor---OPS-SAT VOLT---is going to be launched in late 2025 and will make a great opportunity to validate the algorithms developed based on our benchmark in the wild after deploying them on-board an operational satellite.

\section*{Methods}
 
\subsection*{Data acquisition and annotation}

\begin{figure}[ht]
\centering
\includegraphics[width=.5\linewidth]{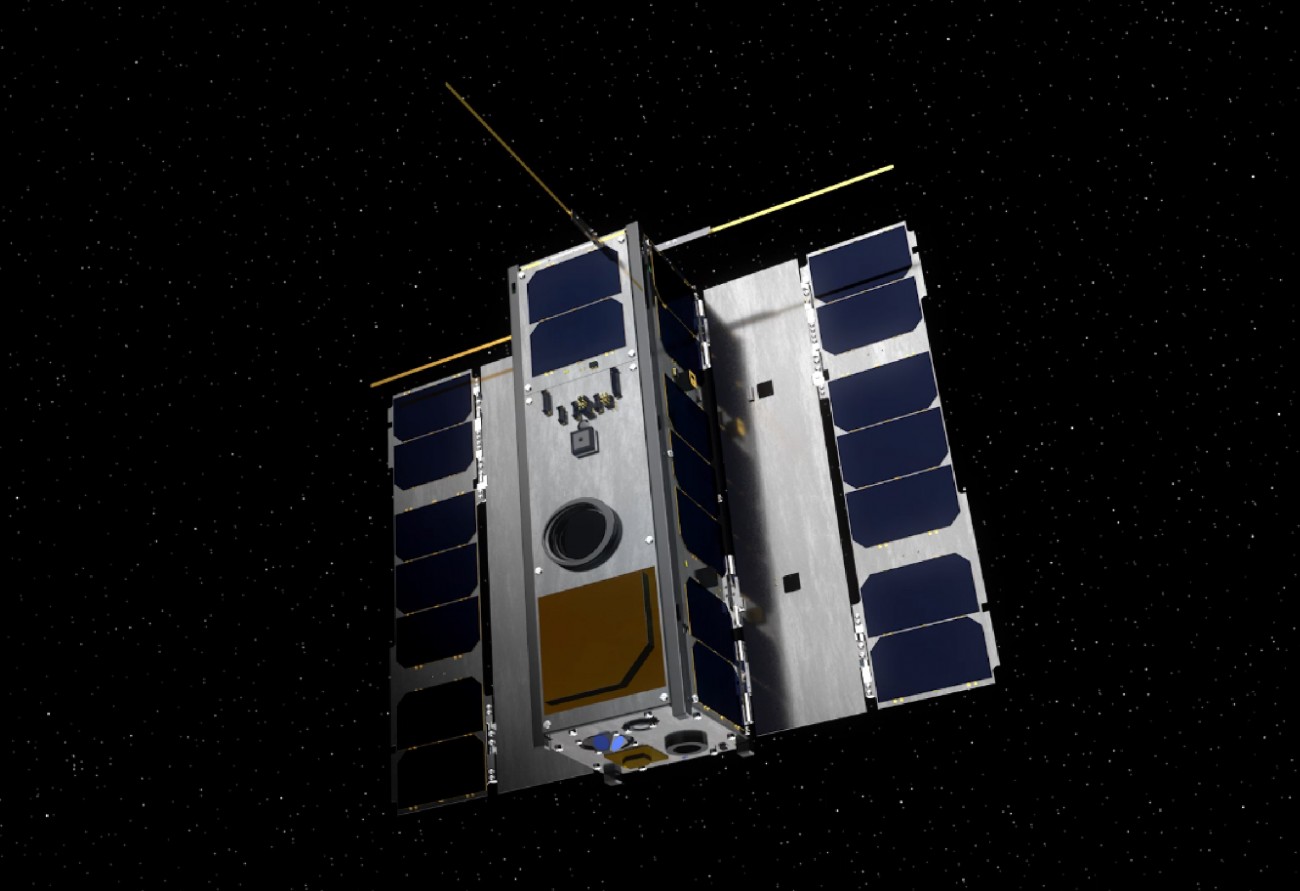}
\caption{The ESA OPS-SAT frontal view. Image credits: European Space Agency.}
\label{fig:aircraft}
\end{figure}

The telemetry data delivered\cite{zenodo24opssatad} in this paper was acquired from the ESA OPS-SAT satellite (Figure~\ref{fig:aircraft}). It is a small 3-unit (3U, where 1U=$10$\,cm$^3$) CubeSat launched in December 2019 with the primary objective of being a technological demonstrator for in-orbit data processing. It finished its mission with the atmospheric reentry on 22 May 2024, but it generated lots of useful data during more than 4 years of its operations, including satellite imagery~\cite{shendy_few-shot_2024} and telemetry~\cite{OPSiccs23}.

\noindent OPS-SAT offered a unique opportunity for researchers to run their experiments and algorithms in orbit. While these experiments were carried out, all telemetry data was simultaneously collected and recorded in the ESA archive. The archive was monitored for potential anomalies to ensure the mission's stable and uninterrupted operation. Our dataset consists of telemetry fragments recommended by the OPS-SAT operation engineers as the most ``interesting'' (according to their subjective assessment) for anomaly detection. The actual data collection process was carried out using the data exchange platform WebMUST\cite{EsaMust21} used in the European Space Operations Centre (ESOC). This platform is restricted to the authorized ESA partners only, but the data included in our dataset package does not have to be requested through it, and thus is made publicly available. 

The online OXI tool for visualization and annotation of satellite telemetry (\url{https://oxi.kplabs.pl/})\cite{oxi23} was used to enable a collaborative labeling process of the dataset. Using this application, domain experts were able to manually extract and annotate telemetry segments representing periods of nominal and anomalous operation. The initial selection of anomalies was provided by 3 ESA spacecraft operations engineers and further curated by 2 machine learning experts (with more than 10 years of experience each). The curated annotations were finally reviewed by the three spacecraft operations engineers. The detailed satellite telemetry annotation process, together with the visual artefacts generated throughout it, are discussed in~\cite{OPSiccs23,NalepaOPSSAT2022,OPSiac23}.

\subsection*{Feature extraction}

Due to the characteristics of satellite telemetry, the segments of raw data selected by the domain experts have varying lengths and sampling frequency. As such, they could not be handled by most machine learning algorithms without performing an additional preprocessing or feature extraction. Thus, 18 handcrafted features were designed for the task of anomaly detection~\cite{OPSiccs23}---they were calculated separately for each segment, and they are included in our benchmark. An algorithm operating on such features was already validated in our previous work focusing on the application of data-driven anomaly detection on board OPS-SAT~\cite{OPSiccs23}. 

\begin{figure}
    \centering
    \includegraphics[width=0.75\linewidth]{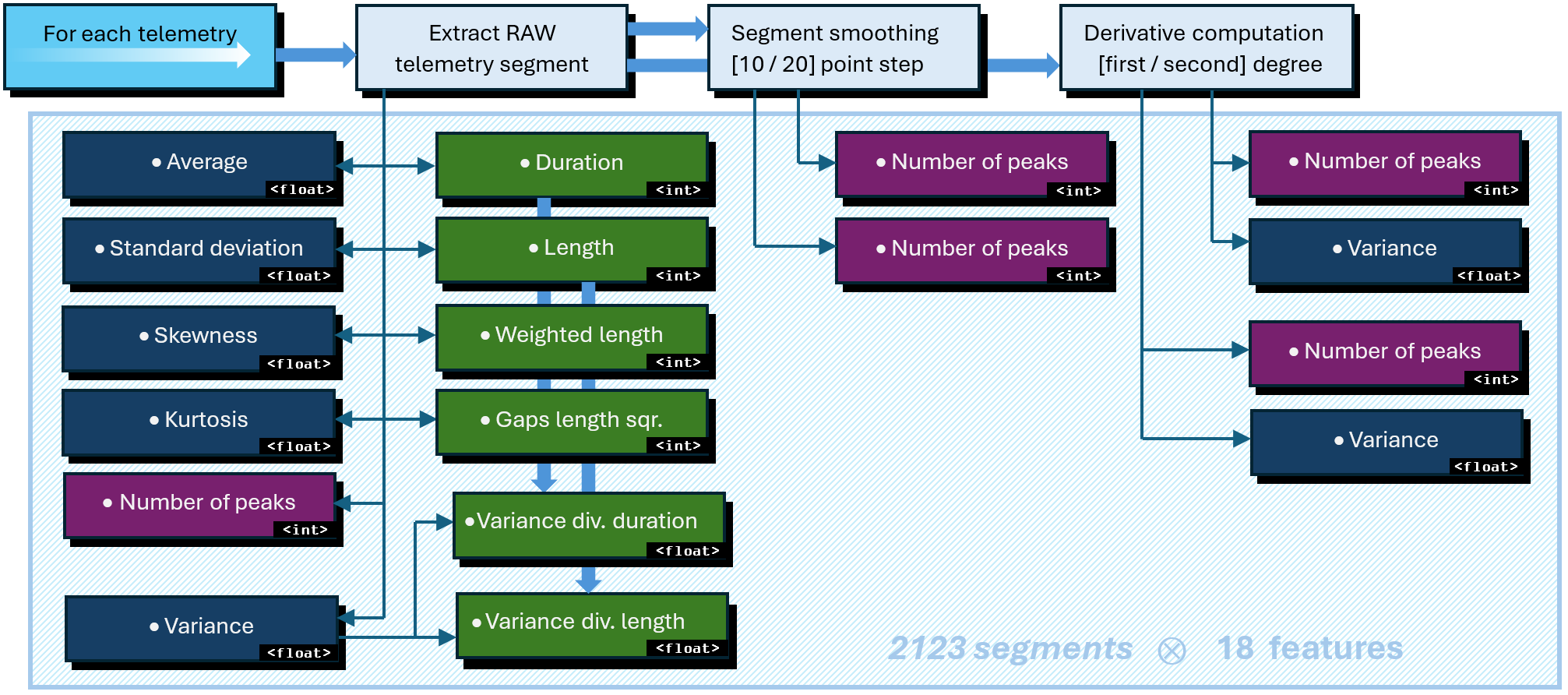}
    \caption{The features extracted for each segment, with the corresponding data type. The meaning of the colors: dark blue for popular statistics, violet for peak counters for various converted segments, and green for the length or duration-related features.}
    \label{fig:datasetfeatures}
\end{figure}

The features extracted for each telemetry segment are presented in Figure~\ref{fig:datasetfeatures}. They are divided into three groups:
\begin{itemize}
    \item \textbf{12 features extracted from raw segments}, including basic statistics, such as the arithmetic average of the signal values, their standard deviation, skewness, kurtosis and variance ($\langle mean \rangle$, $\langle std \rangle$, $\langle skew \rangle$, $\langle kurtosis \rangle$, and $\langle var \rangle$), but also the number of peaks (of the minimum of 10\% prominence, with a peak prominence measuring how much a peak ``stands out'' in relation to the signal, while considering its height and location: $\langle n\_peaks \rangle$), duration (in seconds: $\langle duration \rangle$) and the length (in the number of telemetry points: $\langle len \rangle$), the weighted length (weighted by sampling: $\langle len\_weighted \rangle$), the gaps' length (the squared number of missing data points: $\langle gaps\_squared \rangle$), and the weighted variance (weighted by the duration and by the length: $\langle var\_div\_duration \rangle$, $\langle var\_div\_len \rangle$).
    \item \textbf{2 features extracted from the smoothed segments (using the uniform interpolation\cite{ZHANG21})}, including the number of peaks (extracted using the 10 and 20 points smoothing steps: $\langle smooth10\_n\_peaks \rangle$, $\langle smooth20\_n\_peak \rangle$).
    \item \textbf{4 features extracted from the first and the second derivatives of the segment}, including the number of peaks and variance ($\langle diff\_peaks \rangle$, $\langle diff2\_peaks \rangle$, $\langle diff\_var \rangle$, $\langle diff2\_var \rangle$).
    
\end{itemize}


\noindent Employing the duration, the length and the gaps' length features should allow the algorithms to easily capture some ``obvious'' abnormalities in the telemetry data. This intuitively could lead to promote some less computationally demanding AD methods suitable for on-board applications. The proposed set of features serves as an example and may be easily expanded (or replaced) by the community by (\textit{i})~designing new feature extractors (potentially followed by feature selectors), (\textit{ii})~using other well-established feature sets~\cite{lubba_catch22_2019} or (\textit{iii})~benefiting from the automated feature learning~\cite{Tafazoli2024}.

\subsection*{Benchmarking procedure}

We provide a procedure that should be followed to confront the AD algorithms over our dataset. The entire dataset of 2123 telemetry segments is split into the training ($\TrainingSet$) and test ($\TestSet$) sets (Table~\ref{tab0_dataset}), forming an AI-ready dataset. To extract these subsets, we performed the stratified random sampling to maintain the original percentage of anomalies in both $\TrainingSet$ and $\TestSet$.

\begin{table}[ht!]
\centering
\caption{The number of telemetry segments included in the training ($\TrainingSet$) and test ($\TestSet$) sets.}\label{tab0_dataset}
\renewcommand{\tabcolsep}{2.5mm}
\begin{tabular}{r|cc|c}
\hline
Class & Training set ($\TrainingSet$) & Test set ($\TestSet$) & Total \\
\hline
Nominal & 1273 &  416 & 1689 \\
Anomalous & 321 & 113 & 434 \\
\hline
Sum & 1494 &  529 & 2123 \\
\hline
\end{tabular}
\end{table}


\noindent The benchmarking procedure can be summarized by the following steps:
\begin{enumerate}
    \item Load the dataset from the \texttt{dataset.csv} file.
    \item Split the dataset into $\TrainingSet$ and $\TestSet$ according to the $\langle train \rangle$ attribute included in this file.
    \item \textbf{[Optionally]} Preprocess the datasets using e.g., data normalization, additional feature extraction, feature selection and other steps directly related to the AD algorithm which undergoes the benchmarking process.
    \item \textbf{[Optionally]} Train a machine learning model over $\TrainingSet$.
    \item Quantify the algorithm's performance over $\TestSet$ using the metrics discussed in the next section. 
\end{enumerate}




\subsection*{Quality metrics}\label{subsection:metrics}




The following metrics should always be calculated over the test set $\TestSet$ while confronting the AD algorithms (both supervised and unsupervised) over the dataset OPSSAT-AD introduced in this work:
\begin{itemize}
    \item \textbf{Accuracy:} $({TP+TN}) / ({TP+TN+FP+FN})$,
    \item \textbf{Precision:} $TP / (TP+FP)$,
    
    \item \textbf{Recall:} $TP / (TP+FN)$,
    
    \item \textbf{$F_1$ score:} $(2 \cdot {\rm precision} \cdot {\rm recall}) / ({\rm precision} + {\rm recall})$,
    \item \textbf{Matthews' Correlation Coefficient (MCC)~\cite{Baldi00mmc}:} $(TP \cdot TN - FP \cdot FN ) / \sqrt{(TP+FP)\cdot(TP+FN)\cdot(TN+FP)\cdot(TN+FN)}$,
    \item \textbf{Area under the receiver operating characteristic curve (AUC$_{\rm ROC}$)},
    \item \textbf{Area under the precision-recall curve (AUC$_{\rm PR}$)},
\end{itemize}

\noindent where $TP$, $TN$, $FP$, and $FN$ are the number of true positives (anomalous telemetry segments correctly identified as anomalies), true negatives (nominal telemetry segments correctly identified as nominal), false positives (nominal telemetry segments incorrectly identified as anomalies), and false negatives (anomalous telemetry segments incorrectly identified as nominal). All metrics should be maximized ($\uparrow$), with one indicating the best score (MCC ranges from $-1$ to $1$, other metrics from $0$ to\,$1$).

\subsection*{The baseline: anomaly detection algorithms}\label{subsection:algorithms}

Although there are ground-truth AD datasets that may be used to train supervised models for this task, they are extremely limited and, by definition, they cannot capture a representative set of anomalies (otherwise such ``anomalies'' would not be ``anomalies'' any longer). In practice, while building data-driven AD algorithms for satellite telemetry, practitioners may not be able to access real-life ground-truth data, hence unsupervised methods have been gaining research attention. Here, we establish a set of baseline results obtained using 30 AD methods, including both supervised and unsupervised algorithms (Table~\ref{tab:detectors}).

\begin{table}[ht!]
    \centering
    \begin{tabular}{lc|l}
\hline
   Abbreviation & Year & Algorithm \\
\hline
       \multicolumn{2}{c}{\textbf{\textit{Supervised algorithms}}} \\
\hline

    Linear+L2 \cite{Gruning2006} & 2006 & Linear classifier with $L_2$ regularization \\
    LR \cite{linear2008} & 2008 & Logistic regression \\
    AdaBoost\cite{Hastie09} & 2009 & Adaptive Boosting \\ 
    LSVC \cite{LeeLin13} & 2013 & Support Vector Classifier with the squared hinge linear loss \\
    XGBOD\cite{YueHryniewicki18} & 2018 & Extreme Gradient Boosting Outlier Detection \\
    FCNN\cite{Kwon2019} & 2019 & Fully Connected Neural Network with dropout and batch normalization\\
    RF+ICCS\cite{OPSiccs23} & 2023 & Random Forest based model with segment augmentation \\

\hline
        \multicolumn{2}{c}{\textbf{\textit{Unsupervised algorithms}}} \\
\hline
    PCA\cite{Mastrangelo96,Aggarwal2017} & 1996 & Principal Component Analysis \\
    LMDD\cite{LMDDArning96} & 1996 & Linear Method for Deviation Detection \\
    COF\cite{Tang02} & 2002 & Connectivity-based Outlier Factor\\
    KNN\cite{Angiulli02} & 2002 & K-Nearest Neighbors \\
    CBLOF\cite{HEHU03}  & 2003 & Cluster-Based Local Outlier Factor\\
    ABOD\cite{ABODKriegel08} & 2008 & Angle-based Outlier Detector\\
    IForest\cite{LiuTing08} & 2008 & Isolation Forest \\
    SOD\cite{Kriegel09} & 2009 & Outlier Detection in Axis-Parallel Subspaces of High Dimensional Data\\
    SOS\cite{SOSJanssens12} & 2012 & Stochastic Outlier Selection\\
    VAE \cite{Kingma13} & 2013 & Variational Autoencoder \\
    OCSVM\cite{Erfani16} & 2016 & One-Class Support Vector Machine with a polynomial kernel \\
    LODA\cite{Pevny2016} & 2016 & Lightweight On-line Detector of Anomalies \\
    GMM\cite{Aggarwal2017}  & 2017 & Gaussian Mixture Model \\
    AnoGAN \cite{Schlegl17} & 2017 & Generative Adversarial Networks for AD \\
    DeepSVDD\cite{Ruff18a} & 2018 & Deep one-class classification \\
    ALAD \cite{Houssam18} & 2018 & Generative Adversarial Networks for AD \\
    INNE\cite{InneBandaragoda18} & 2018 & Isolation-based Anomaly Detection Using Nearest-Neighbor Ensembles \\
    SO-GAAL\cite{Liu20} & 2020 & Single-objective Generative Adversarial Active Learning \\
    MO-GAAL\cite{Liu20} & 2020 & Multi-objective Generative Adversarial Active Learning \\    
    COPOD \cite{Li2020COPODCO} & 2020 & Copula-based outlier detection \\
    ECOD\cite{LiZheng2022} & 2022 & Empirical Cumulative Distribution Functions \\
    LUNAR \cite{Goodge_Hooi_Ng_Ng_2022} & 2022 & Unified Local Outlier Detection with Graph Neural Networks\\
    DIF \cite{XuPang23} & 2023 & Deep Isolation Forest \\
\hline
       
    \end{tabular}
    \caption{Anomaly detection methods investigated in this study.}
    \label{tab:detectors}

\end{table}

For all those algorithms, implemented in the PyOD framework~\cite{zhao2019pyod} (\url{https://pyod.readthedocs.io/en/latest/}), the default parameters (suggested by the authors of these techniques) are used, with the anomaly contamination factor set to 0.2, according to the anomaly distribution observed in $\TrainingSet$. To ensure reproducibility, we provide a Jupyter Notebook showing how to execute an example AD algorithm, in a both supervised and unsupervised training regime (\textit{modeling\_examples.ipynb}).

\section*{Dataset Layout}

\begin{figure}[ht!]
         \centering
         \includegraphics[width=1.8in]{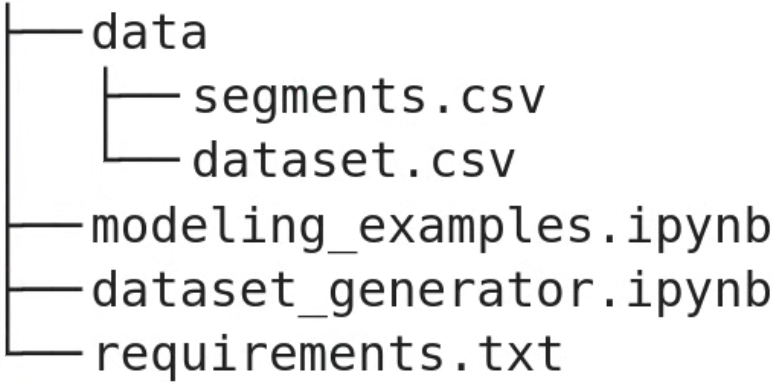}
         \caption{The layout of the OPS-SAT benchmark for anomaly detection.}
         \label{fig:package}
    \end{figure}

The dataset\cite{zenodo24opssatad} is built of 9 source telemetry channels that were selected by the space operations engineers. They include 3 magnetometer telemetry channels: I\_B\_FB\_MM\_0 (CADC0872), I\_B\_FB\_MM\_1 (CADC0873), I\_B\_FB\_MM\_2 (CADC0874), and 6 photo diode (PD) channels: I\_PD1\_THETA (CADC0884), I\_PD2\_THETA (CADC0886), I\_PD3\_THETA (CADC0888), I\_PD4\_THETA (CADC0890), I\_PD5\_THETA (CADC0892), I\_PD6\_THETA (CADC0894). Here, the names correspond to the source names from the WebMUST repository and the OPS-SAT telemetry channel names (in brackets). The layout of the dataset is summarized in Figure~\ref{fig:package}---it includes both the raw files, as well as the extracted features in a tabular form.

\subsection*{Raw telemetry data}

\begin{figure}[ht!]
        \centering
        \includegraphics[width=.65\linewidth]{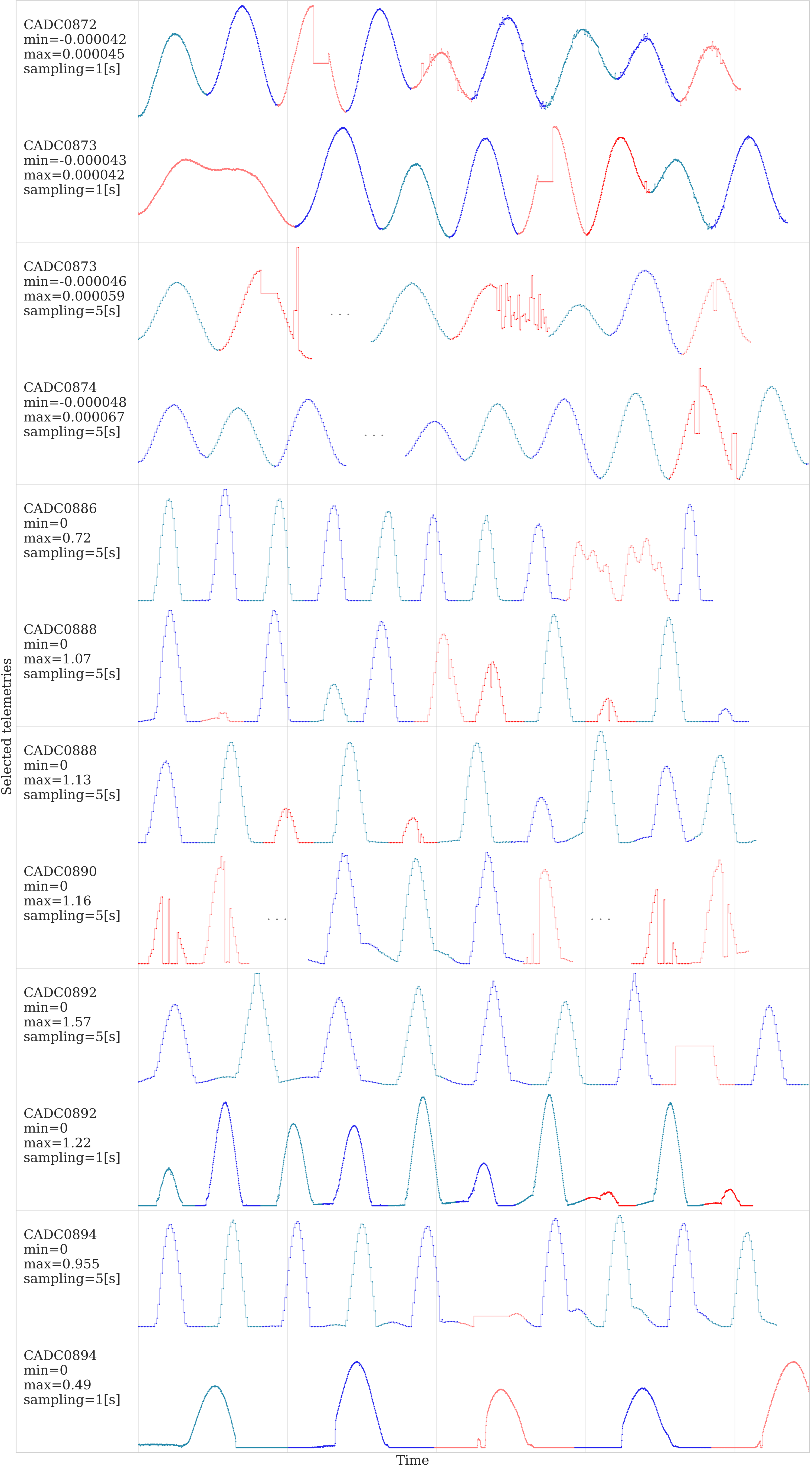}
        \caption{Selected segments from out OPS-SAT dataset. Several types of signal distortions are depicted, including peaks, deformations, noise (CADC0873), irregular periodicity (CADC0886), short (CADC0892, CADC0894) and long data gaps (CADC0874). Anomalous segments are plotted in red. For brevity, we omit the axis values, but provide data ranges and the sampling information for each channel.}
        \label{fig:example_segments}
    \end{figure}

In Figure~\ref{fig:example_segments}, we visualize selected characteristics of the acquired telemetry signals, effectively showing real-world challenges concerned with telemetry data acquired in the wild (e.g., missing readouts, different sampling frequencies). Such segments for all the aforementioned telemetry channels and their selected parts are included in the \texttt{data/segments.csv} file. It contains the attributes that identify the registration time: $\langle timestamp\rangle$ (ISO date format), $\langle channel\rangle$ (the channel name), $\langle value\rangle$ (the acquired signal value), and $\langle label\rangle$ (the ground-truth annotation). Additionally, we provide the consecutive segment numbers ($\langle segment\rangle$), their sampling rate ($\langle sampling\rangle$), and the indication if they are included in $\TrainingSet$ ($\langle train\rangle$).


    
\subsection*{Extracted features}

In the tabular version of the dataset (\texttt{data/dataset.csv}), we include the extracted features. In the Supplementary Materials, we present the distributions of all of the provided features, rendered for both $\TrainingSet$ and $\TestSet$ sets of our dataset (Figure~\ref{fig:features}).
    
\section*{Experimental Validation}

\begin{figure}[ht!]
\centering
\includegraphics[width=.52\linewidth]{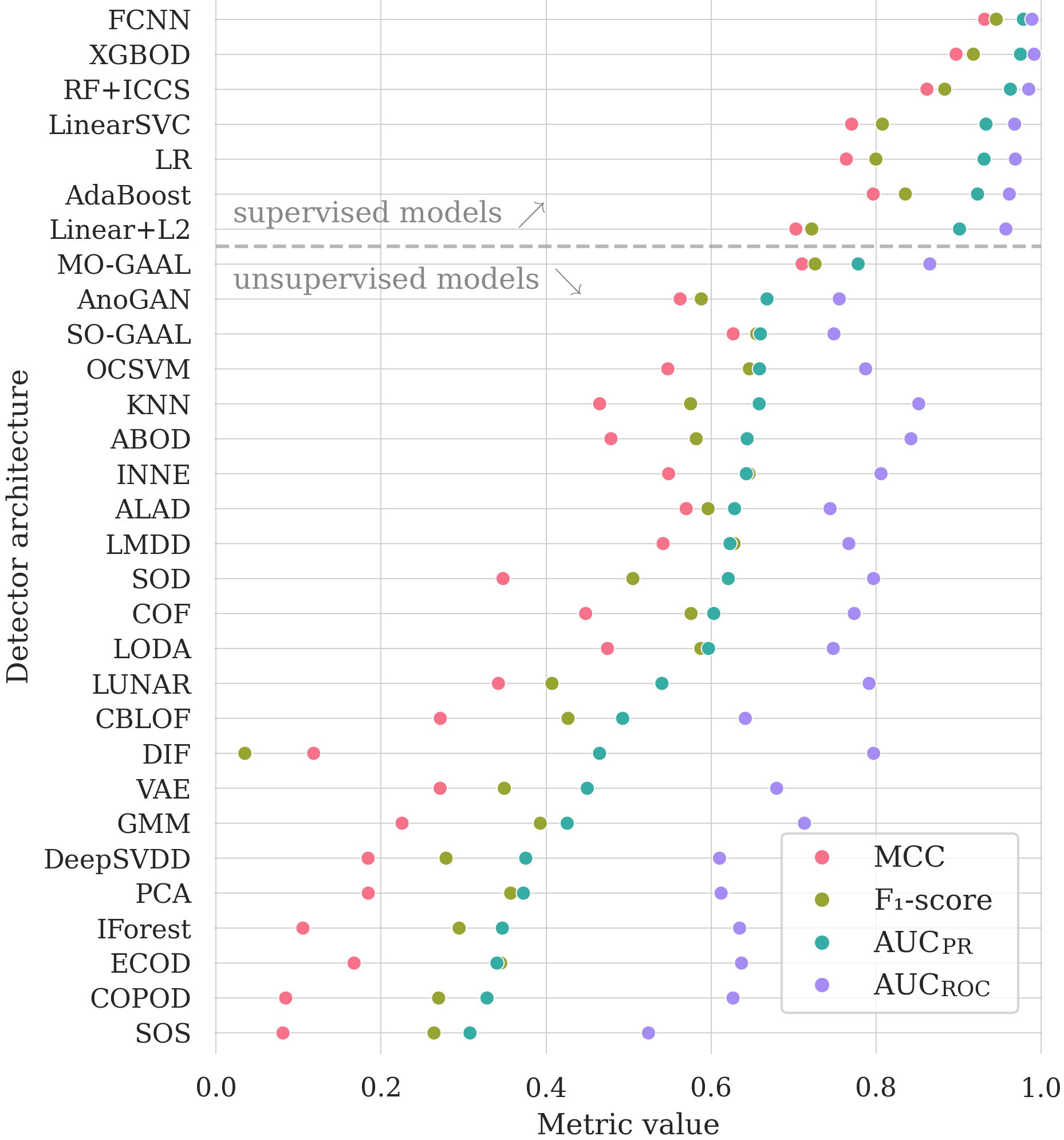}
\caption{The results (over $\TestSet$) obtained using the investigated machine learning models (first grouped according to their training strategy, either supervised or unsupervised, than sorted by $F_1$).}
\label{fig:metrics}
\end{figure}
\begin{figure}[ht!]
\centering
\includegraphics[width=.52\linewidth]{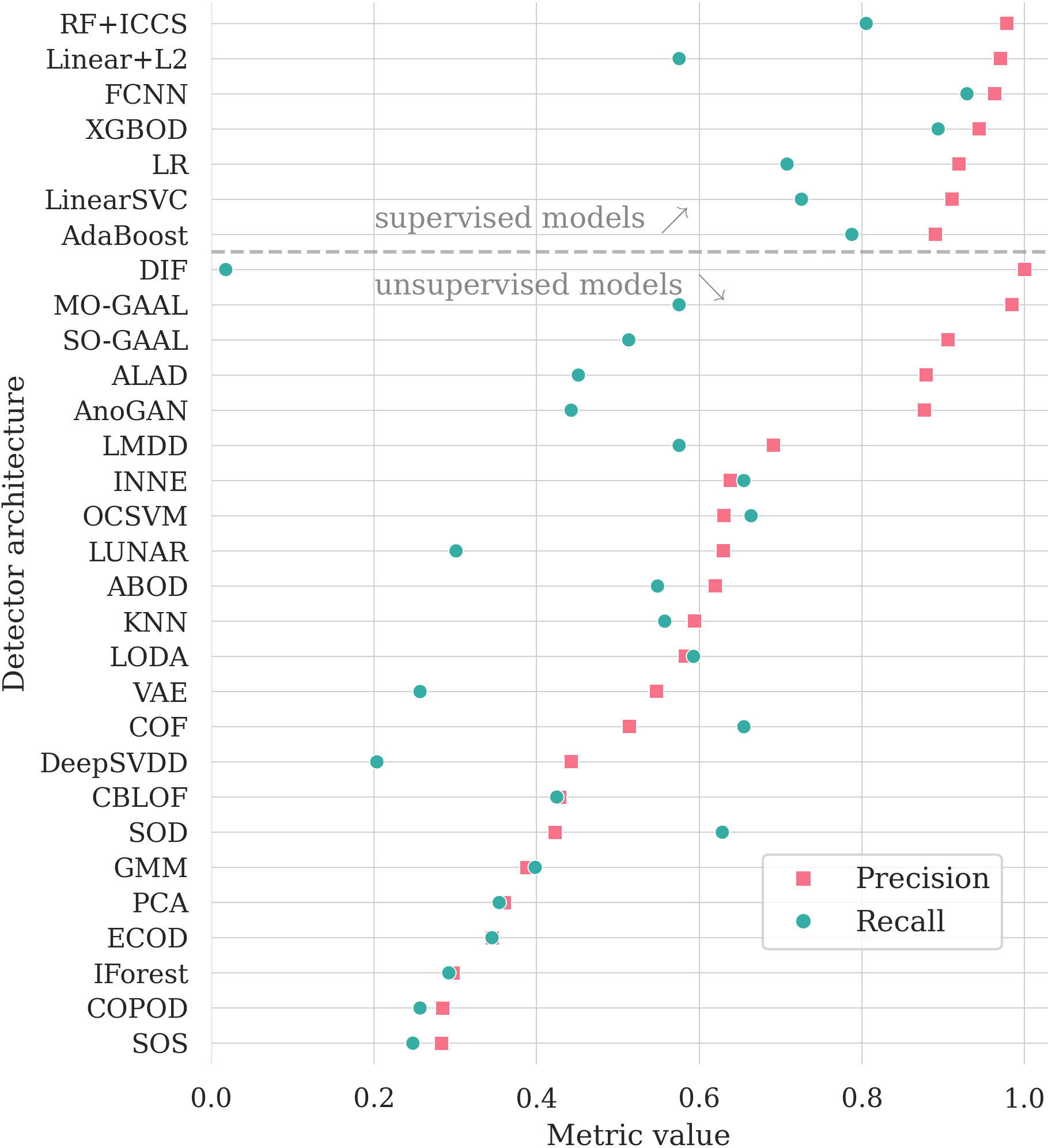}
\caption{Precision and recall metrics (over $\TestSet$) obtained using the investigated algorithms. Models are sorted according to the number of misclassified telemetry segments, with the best-performing one rendered on top of the graph.}
\label{fig:prec_rec}
\end{figure}

In Table~\ref{tab2_results}, we aggregate the results obtained using all the investigated AD algorithms. Here, we highlighted the globally best results (in bold for each quality metric), and we underlined the best results elaborated by the unsupervised algorithms, as we consider them a different category of the AD solutions. We are aware that some algorithms from PyOD should be rather trained using nominal data only (i.e., OCSVM or autoencoders) to achieve better results. As an example, the OCSVM model achieves $AUC_{PR}$ of 0.659  and $AUC_{ROC}$ of 0.787 in our setting, but when using only the nominal data (without abnormal segments) for training, the corresponding values are 0.762 and 0.815. However, we wanted our baseline to be consistent and to reflect a typical usage of the PyOD framework by a non-expert user. Also, the fine-tuning of those algorithms is out of the scope of this study. In Figure~\ref{fig:metrics}, we render the selected metrics for each model. We can indeed observe the better performance of supervised methods, as those could actively benefit from the labeled anomaly examples while building a machine learning model. In Figure~\ref{fig:prec_rec}, we also display the precision and recall  quality metrics. For the fully-connected neural network, we can observe only four false positives and eight false negatives of all $\TestSet$ samples, reaching the precision of 0.963, and the recall of 0.929.



\begin{table}[ht!]
\centering
\renewcommand{\tabcolsep}{2.5mm}    
\begin{tabular}{l|cc|ccccc}
\hline    
Model & $AUC_{PR}$ ($\uparrow$) & $AUC_{ROC}$ ($\uparrow$) & $Accuracy$ ($\uparrow$) & $F_1$ ($\uparrow$) & $Precision$ ($\uparrow$) & $Recall$ ($\uparrow$) & $MCC$ ($\uparrow$) 
\\
\hline
\multicolumn{2}{c}{\textbf{\textit{Supervised algorithms}}} \\
\hline

FCNN & \textbf{0.979} & 0.989 & \textbf{0.977} & \textbf{0.946} & 0.963 & \textbf{0.929} & \textbf{0.932} \\ 
XGBOD & 0.975 & \textbf{0.992} & 0.966 & 0.918 & 0.944 & 0.894 & 0.897 \\ 
RF+ICCS & 0.963 & 0.985 & 0.955 & 0.883 & 0.978 & 0.805 & 0.862 \\ 
LSVC & 0.934 & 0.968 & 0.926 & 0.808 & 0.911 & 0.726 & 0.771 \\ 
LR & 0.931 & 0.969 & 0.924 & 0.800 & 0.920 & 0.708 & 0.764 \\ 
AdaBoost & 0.923 & 0.962 & 0.934 & 0.836 & 0.890 & 0.788 & 0.797 \\ 
Linear+L2 & 0.901 & 0.958 & 0.905 & 0.722 & 0.970 & 0.575 & 0.703 \\ 
\hline
\multicolumn{2}{c}{\textbf{\textit{Unsupervised algorithms}}} \\
\hline
MO-GAAL & \underline{0.779} & \underline{0.865} & \underline{0.907} & \underline{0.726} & 0.985 & 0.575 & \underline{0.710} \\ 
AnoGAN & 0.668 & 0.756 & 0.868 & 0.588 & 0.877 & 0.442 & 0.563 \\ 
SO-GAAL & 0.660 & 0.749 & 0.885 & 0.655 & 0.906 & 0.513 & 0.627 \\ 
OCSVM & 0.659 & 0.787 & 0.845 & 0.647 & 0.630 & \underline{0.664} & 0.548 \\ 
KNN & 0.658 & 0.852 & 0.824 & 0.575 & 0.594 & 0.558 & 0.465 \\ 
ABOD & 0.644 & 0.843 & 0.832 & 0.582 & 0.620 & 0.549 & 0.479 \\ 
INNE & 0.643 & 0.806 & 0.847 & 0.646 & 0.638 & 0.655 & 0.549 \\ 
ALAD & 0.629 & 0.744 & 0.870 & 0.596 & 0.879 & 0.451 & 0.570 \\ 
LMDD & 0.623 & 0.767 & 0.854 & 0.628 & 0.691 & 0.575 & 0.542 \\ 
SOD & 0.621 & 0.797 & 0.737 & 0.505 & 0.423 & 0.628 & 0.348 \\ 
COF & 0.603 & 0.774 & 0.794 & 0.576 & 0.514 & 0.655 & 0.448 \\ 
LODA & 0.597 & 0.748 & 0.822 & 0.588 & 0.583 & 0.593 & 0.475 \\ 
LUNAR & 0.540 & 0.792 & 0.813 & 0.407 & 0.630 & 0.301 & 0.342 \\ 
CBLOF & 0.493 & 0.642 & 0.756 & 0.427 & 0.429 & 0.425 & 0.272 \\ 
DIF & 0.465 & 0.797 & 0.790 & 0.035 & \underline{\textbf{1.000}} & 0.018 & 0.118 \\ 
VAE & 0.450 & 0.680 & 0.796 & 0.349 & 0.547 & 0.257 & 0.272 \\ 
GMM & 0.426 & 0.713 & 0.737 & 0.393 & 0.388 & 0.398 & 0.225 \\ 
DeepSVDD & 0.375 & 0.610 & 0.775 & 0.279 & 0.442 & 0.204 & 0.184 \\ 
PCA & 0.373 & 0.612 & 0.728 & 0.357 & 0.360 & 0.354 & 0.185 \\ 
IForest & 0.347 & 0.635 & 0.701 & 0.295 & 0.297 & 0.292 & 0.105 \\ 
ECOD & 0.340 & 0.637 & 0.720 & 0.345 & 0.345 & 0.345 & 0.167 \\ 
COPOD & 0.328 & 0.627 & 0.703 & 0.270 & 0.284 & 0.257 & 0.084 \\ 
SOS & 0.308 & 0.524 & 0.705 & 0.264 & 0.283 & 0.248 & 0.081 \\ 
\hline

\end{tabular}
\caption{The experimental results, sorted by $AUC_{PR}$. The globally best results for each metric are \textbf{boldfaced}, and the best among the unsupervised algorithms are \underline{underlined}.}\label{tab2_results}
\end{table}



The investigation of the unsupervised algorithms reveals that some of them reach a point where they return a small set of mistakenly assessed telemetry samples. Especially the detectors built upon MO-GAAL, SO-GAAL and AnoGAN offered high precision. In terms of the number of misclassified examples, MO-GAAL obtained a better result than the supervised methods, and made one less false detection when compared to FCNN. A number of other unsupervised algorithms, however, tend to either return a large number of false negatives, or to raise many false alarms with a low number of false negatives. In the first group of such methods, we can observe DIF, ALAD, DeepSVDD, and AnoGAN, whereas e.g.,~COF belongs to the second group here. The usability of such algorithms would be rather limited to situations when avoiding one type of the classification error could be more practically important (e.g., to minimize the overhead induced on the space operations teams that would have to review many incorrectly raised false alarms). 

\section*{Usage Notes}


The dataset\cite{zenodo24opssatad} contains the data in two different forms: a set of the original telemetry segments and a corresponding set of handcrafted features, both with anomaly labels. Both collections are also encoded in the popular, easy-to-handle CSV format and are ready to to use with various machine learning models (thus, they can be considered AI-ready). All the algorithms were implemented in \texttt{Python} using \texttt{PyOD}\cite{pyod2019zhao} 1.1.2, \texttt{TensorFlow}\cite{tensorflow2015-whitepaper} 2.15, and \texttt{PyTorch}\cite{pytorch2017paszke-automatic} 2.1.2. Additionally, we used \textit{NumPy}\cite{numpy2020array-harris} 1.26.2 and \textit{Pandas}\cite{2020pandas} 2.1.14 for data preparation, \textit{Seaborn}\cite{seaborn2021Waskom} 0.13.0 for the visualizations, as well as \textit{OXI}\cite{oxi23} for the initial data analysis and labeling processes. Finally, our benchmark is accompanied with a {Jupyter Notebook}, containing an example experiment (\texttt{modeling\_examples.ipynb}), in order to ensure the experimental reproducibility.

\section*{Code Availability}


The code for working with the OPS-SAT benchmark, including the functionalities used to prepare the numerical results, figures, and tables for this article, is available through the following GitHub repository: \url{https://github.com/kplabs-pl/OPS-SAT-AD} under the MIT license.



\bibliography{main}

\section*{Acknowledgements}

This work was partially supported through the following projects: 
“On-board Anomaly detection from the OPS-SAT telemetry using deep learning” (40001373339/22/NL/GLC/ov) and "Few-shot anomaly detection in satellite telemetry" (4000141301) funded by the European Space Agency. JN was supported by the Silesian University of Technology grant for maintaining and developing research potential.

\section*{Author contributions statement}


B.R., K.K. and J.N. conceived and designed the study. 
B.R. and D.E. acquired the data and performed the experiments.
B.R. implemented the computational pipeline and performed the analyses. 
B.R. drafted the manuscript. K.K. and J.N. edited and improved the manuscript. 
D.E. revised the manuscript. All authors have read and approved the final manuscript.





\section*{Corresponding author}
Correspondence to \href{mailto:b.ruszczak@po.edu.pl}{Bogdan Ruszczak}.

\section*{Competing interests}

The authors declare no competing interests.

\section*{Supplementary Materials}

\begin{figure}[ht!]
    \centering
    \includegraphics[width=1.0\linewidth]{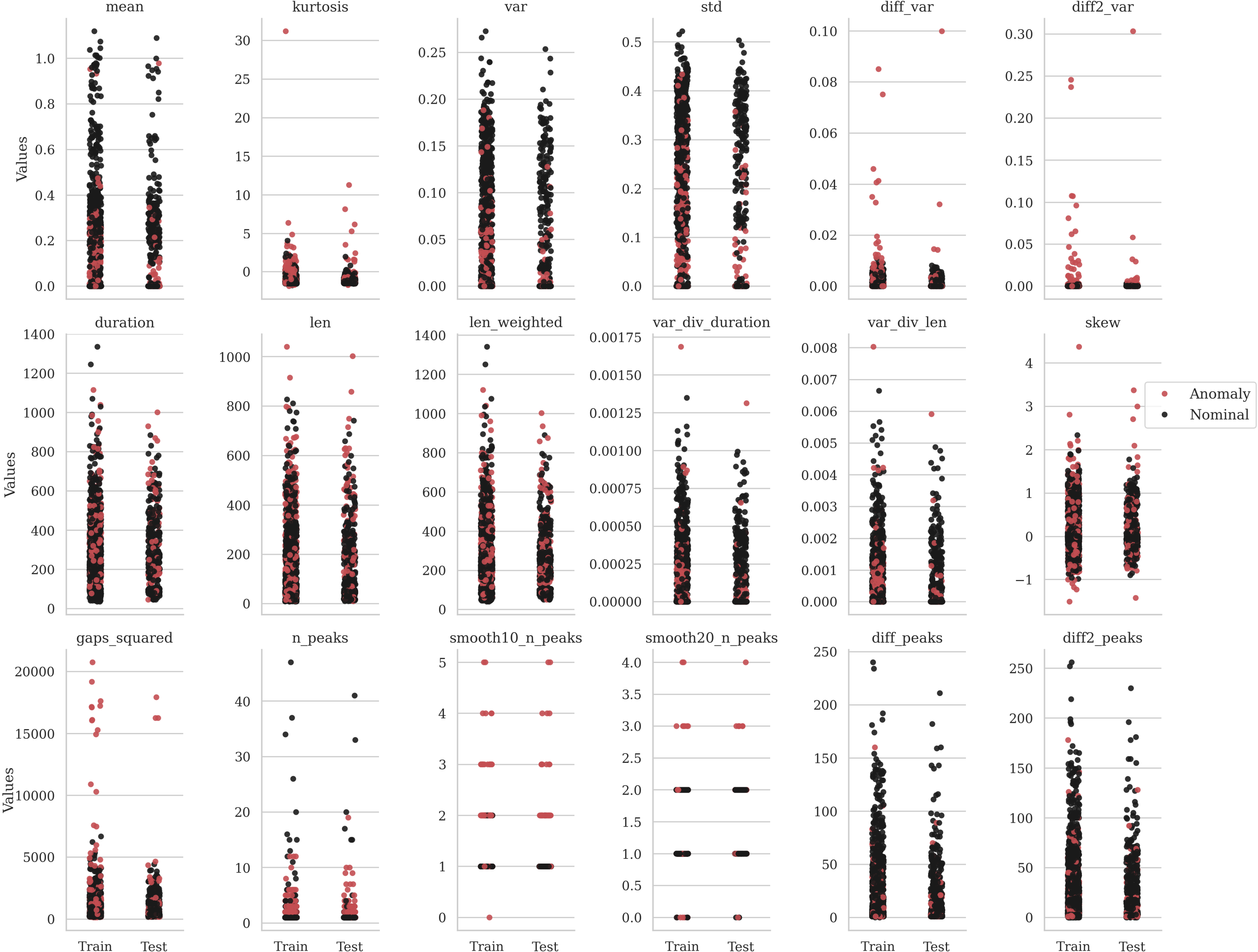}
    \caption{Dataset features for the training and validation subsets with the indication of anomalies (marked in red).}
    \label{fig:features}
\end{figure}

\begin{figure}[ht]
\centering
\includegraphics[width=.8\linewidth]{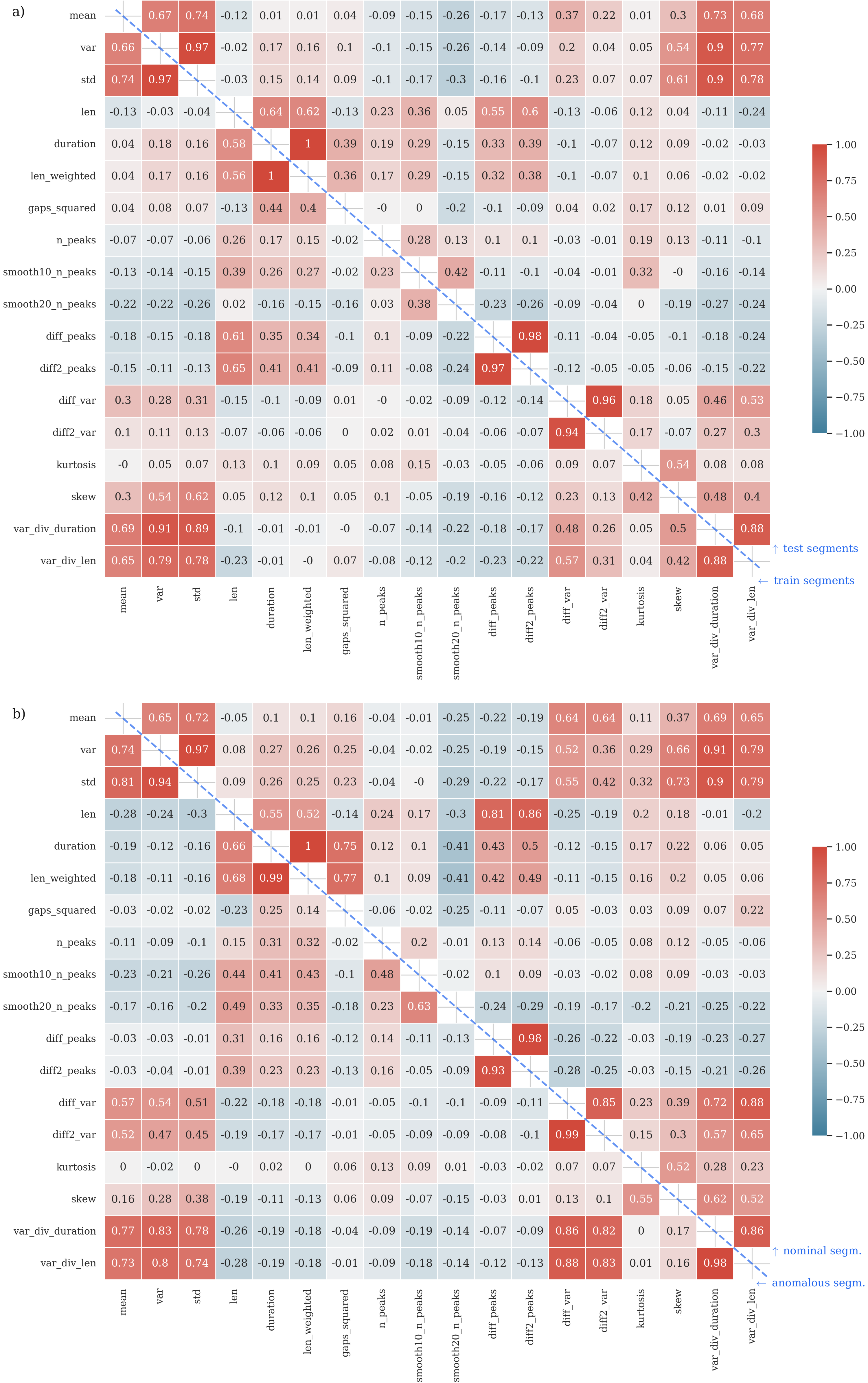}
\caption{The coefficient of correlation computed between each feature, for the training (below the blue dashed diagonal line) and test set (above the same diagonal line). We employed: (a) Pearson's correlation coefficient and (b)~Spearman's Rank correlation coefficient.}
\label{fig:corrs1}
\end{figure}

The distribution of the extracted features elaborated for each telemetry segment included in our dataset is depicted in Figure~\ref{fig:features}. We compare the distribution for the training ($\TrainingSet$) and test ($\TestSet$) sets, to visualize the effect of the training-test dataset split on the feature distributions. Figure~\ref{fig:corrs1} provides a detailed view of the relations between the extracted features for the $\TrainingSet$ and $\TestSet$ sets. We rendered this plot to confirm that both subsets represent similar data distributions.

\end{document}